\documentclass[aps,prl,twocolumn,superscriptaddress,showpacs,preprintnumbers,amsmath,amssymb]{revtex4}
\usepackage{epsfig}

\newcommand{\piz}{\pi^0}

\newcommand{\etac}{\eta_c}
\newcommand{\etacp}{\eta_c^{\prime}}
\newcommand{\psp}{\psi(3686)}

\newcommand{\jpsi}{J/\psi}
\newcommand{\EE}{e^+e^-}

\newcommand{\pp}{\pi^+\pi^-}
\newcommand{\kk}{K^+K^-}
\newcommand{\ks}{K_S}

\newcommand{\kskp}{K_SK^+\pi^-}
\newcommand{\kkpiz}{\kk\piz}
\newcommand{\kskppp}{K_SK^+\pp\pi^-}
\newcommand{\kkpppiz}{\kk\pp\piz}

\newcommand{\pppppp}{3(\pp)}

\newcommand{\etapp}{\eta\pp}

\newcommand{\ppjpsi}{\pi^+\pi^- J/\psi}

\newcommand{\beq}{\begin{equation}}
\newcommand{\eeq}{\end{equation}}
\newcommand{\bitm}{\begin{itemize}}
\newcommand{\eitm}{\end{itemize}}
\newcommand{\mev}{\mathrm{MeV}}

\newcommand{\mevcc}{\mathrm{MeV}/c^2}
\newcommand{\gev}{\mathrm{GeV}}
\newcommand{\gevc}{\mathrm{GeV}/c}
\newcommand{\gevcc}{\mathrm{GeV}/c^2}

\begin{document}

\preprint{} \preprint{\vbox{ \hbox{   }
                        \hbox{Ver4.2} }}
\title{\quad\\[1.0cm]
Measurements of the mass and width of the $\eta_c$ using $\psi(3686) \to \gamma \eta_c$}

\author{ 
\small
 M.~Ablikim$^{1}$, M.~N.~Achasov$^{5}$, D.~Alberto$^{41}$,
 D.J.~Ambrose$^{38}$, F.~F.~An$^{1}$, Q.~An$^{39}$, Z.~H.~An$^{1}$, J.~Z.~Bai$^{1}$,
 R.~B. Ferroli$^{17}$, Y.~Ban$^{25}$
% ...
\\
\vspace{0.2cm} 
(BESIII Collaboration)\\
\vspace{0.2cm} {\it
$^{1}$ Institute of High Energy Physics, Beijing 100049, People's Republic of China\\
% ...
$^{5}$ G.I. Budker Institute of Nuclear Physics SB RAS (BINP), Novosibirsk 630090, Russia\\
% ...
$^{17}$ INFN Laboratori Nazionali di Frascati , Frascati, Italy\\
% ...
$^{25}$ Peking University, Beijing 100871, People's Republic of China\\
% ...
$^{38}$ University of Rochester, Rochester, New York 14627, USA\\
$^{39}$ University of Science and Technology of China, Hefei 230026, People's Republic of China\\
% ...
$^{41}$ University of Turin and INFN, Turin, Italy\\
% ...
}
\vspace{0.4cm}}


\begin{abstract}
  
The mass and width of the lowest-lying $S$-wave spin singlet charmonium state, the
$\eta_c$, are measured using a data sample of $1.06\times10^8$ $\psi(3686)$ decays
collected with the BESIII detector at the BEPCII storage ring. We use a model that
incorporates full interference between the signal reaction, $\psi(3686)\to\gamma\eta_c$,
and a non-resonant radiative background to describe the line shape of the $\eta_c$
successfully. We measure the $\eta_c$ mass to be $2984.3\pm 0.6\pm 0.6~ \mathrm{MeV}/c^2$
and the total width to be $32.0\pm 1.2\pm 1.0~ \mathrm{MeV}$, where the first errors are
statistical and the second are systematic.

\end{abstract}


\maketitle

In recent years, many new charmonium or charmonium-like states have been discovered. These
states have led to a revived interest in improving the quark-model picture of
hadrons~\cite{many}.  Even with these new discoveries, the mass and width of the
lowest-lying charmonium state, the $\etac$, continue to have large uncertainties when
compared to those of other charmonium states~\cite{PDG}. Early measurements of the
properties of the $\etac$ using $\jpsi$ radiative
transitions~\cite{Baltrusaitis:1985mr,Bai:2003et} found a mass and width near $2978~\mevcc$
and $10~\mev$, respectively. However, recent experiments, including photon-photon fusion
and $B$ decays, have reported a significantly higher mass and a much larger
width~\cite{Asner:2003wv, Aubert:2003pt, Uehara:2007vb, belle2011}. The most recent study
by the CLEO-c experiment~\cite{recent:2008fb}, using both $\psp \to \gamma\etac$ and
$\jpsi\to \gamma\etac$, pointed out a distortion of the $\etac$ line shape in $\psp$
decays. CLEO-c attributed the $\etac$ line-shape distortion to the energy dependence of
the $M1$ transition matrix element.


In this Letter, we report measurements of the $\etac$ mass and width using the radiative
transition $\psp \to \gamma \etac$.  We successfully describe the
measured $\etac$ line shapes using a combination of the energy
dependence of the hindered-$M1$ transition matrix element and a full
interference with non-resonant $\psp$ radiative decays. The analysis is based on a $\psp$
data sample of $1.06\times 10^8$ events~\cite{npsp} collected with the BESIII detector
operating at the BEPCII $\EE$ collider. A 42~pb$^{-1}$ continuum data sample, taken at a
center-of-mass energy of $3.65~\gev$, is used to measure non-$\psp$ backgrounds.

The $\etac$ mass and width are determined from fits to the invariant mass
spectra of exclusive $\etac$ decay modes. Six modes are used to
reconstruct the $\etac$: $\kskp$, $\kkpiz$, $\etapp$,  $\kskppp$,
$\kkpppiz$, and $\pppppp$, where the $\ks$ is reconstructed in
$\pp$, and the $\eta$ and $\piz$ in $\gamma\gamma$ decays. The
inclusion of charge conjugate modes is implied.

The BESIII detector is described in detail in Ref.~\cite{bes3}. The 
detector has a geometrical acceptance of 93\% of $4\pi$. A small cell
helium-based main drift chamber (MDC) provides momentum
measurements of charged particles; in a 1~T magnetic field the
resolution is 0.5\% at $1~\gevc$. It also supplies an energy
loss ($-dE/dx$) measurement with a resolution better than 6\%
for electrons from Bhabha scattering. The electromagnetic calorimeter
(EMC) measures photon energies with a resolution of 2.5\% (5\%) at
$1~\gev$ in the barrel (endcaps). The
time-of-flight system (TOF) is composed of plastic scintillators with
a time resolution of 80~ps (110~ps) in the barrel (endcap) and is mainly
useful for particle identification. The muon system provides $2$~cm
position resolution and measures muon tracks with momenta greater
than $0.5~\gevc$.

We use inclusive Monte Carlo (MC) simulated events as an aid in our background studies.
The $\psp$ resonance is produced by the event generator KKMC~\cite{KKMC}, while the decays
are generated by EvtGen~\cite{EvtGen} with known branching
fractions~\cite{PDG}, or by Lundcharm~\cite{Chen:2000tv} for
unmeasured decays. The signal is generated with an angular
distribution of $1+\cos^{2}\theta_\gamma$ for $\psp\to \gamma \etac$,
and phase space for multi-body $\etac$ decays, where $\theta_\gamma$
is the angle between the photon and the positron beam direction in the
center-of-mass system. Simulated events are processed using
GEANT4~\cite{geant4}, where measured detector resolutions are
incorporated.


We require that each charged track (except those from $\ks$
decays) is consistent with originating from within 1~cm in the
radial direction and 10~cm along the beam direction of the
run-by-run-determined interaction point. The tracks must be within
the MDC fiducial volume, $|\cos\theta|<0.93$. Information from the TOF
and $-dE/dx$ is combined to form a likelihood $\mathcal{L}_\pi$ (or
$\mathcal{L}_K$) for a pion (or kaon) hypothesis. To identify a track as
a pion (kaon), the likelihood $\mathcal{L_\pi}$ ($\mathcal{L_K}$)
is required to be greater than $0.1\%$ and greater than 
$\mathcal{L_K}$ ($\mathcal{L_\pi}$).

Photons are reconstructed from isolated showers in the EMC that are at least 20 degrees
 away from charged tracks. The energy deposited in the nearby TOF scintillator is included
 to improve the reconstruction efficiency and the energy resolution. Photon energies are
 required to be greater than $25~\mev$ in the fiducial EMC barrel region
 ($|\cos\theta|<0.8$) and $50~\mev$ in the endcap ($0.86<|\cos\theta|<0.92$). The showers
 close to the boundary are poorly reconstructed and excluded from the analysis. Moreover,
 the EMC timing, with respect to the collision, of the photon candidate must be in
 coincidence with collision events, {\it i.e.} $0 \leq t \leq 700$ ns, to suppress
 electronic noise and energy deposits unrelated to the event.

The $\ks\to\pi^+\pi^-$ candidates are reconstructed from pairs of oppositely
charged tracks. The secondary vertex constrained tracks must have
an invariant mass $\pm 10~\mevcc$ of the nominal $\ks$
mass, and a decay length more than twice the vertex resolution. The track
information at the secondary vertex is an input to the kinematic fit.
Candidate $\piz$ and $\eta$ mesons are reconstructed from pairs of
photons with an invariant mass in the range $0.118~\gevcc
< M(\gamma\gamma) < 0.150~\gevcc$ for $\piz$ and
$0.50~\gevcc < M(\gamma\gamma) < 0.58~\gevcc$ for
$\eta$. The remaining photons are considered as candidates of the transition photon.

Events with either extra charged tracks or non-zero net charge are rejected.  The $\etac$
candidates are reconstructed from $\kskp$, $\kkpiz$, $\etapp$, $\kskppp$, $\kkpppiz$ and
$\pppppp$. We select events in the region $2.7~\gevcc < M(\etac)<3.2~\gevcc$. 
A four-constraint (4C) kinematic fit of all selected charged
particles and the transition photon with respect to the initial $\psp$
four-momentum is performed to reduce background and improve the mass resolution.
When additional photons are found in an event, we loop over all
possible combinations and keep the one with the best $\chi^2_{\rm 4C}$ from the kinematic
fit. The $\chi^2_{\rm 4C}$ is required to be less than 60, a value is determined by
optimizing the figure of merit for most of the channels, $S/\sqrt{S+B}$, where $S$ ($B$)
is the number of signal (background) events in the signal region ($2.9~\gevcc < M(\etac)
<3.05~\gevcc$). In addition, to remove $\psp\to \ppjpsi$ events, we require there be no
$\pp$ pair with a recoil mass in the $\jpsi$ signal region. To suppress background from
$\piz\to \gamma\gamma$, we demand that the transition photon should not form a $\piz$ with any other
photon in the event.

The main source of background is from $\psp \to \piz X_i$ decays,
where a photon from the $\piz\to \gamma\gamma$ decay is missing, and $X_i$ represents the
$\etac$ final states under study.  These decays could proceed via
various intermediate states, and most of the branching fractions are
unknown. To estimate their contribution, we reconstruct $\psp \to \piz
X_i$ decays from data.  The selection criteria are similar to those
applied to the $\gamma\etac$ candidates except an additional photon
is required.  The $\psp \to \piz X_i$ signal yields are extracted from
fits to the $M(\gamma\gamma)$ invariant mass distributions for
different $M(X_i)$ mass bins. The relative efficiencies of the $\gamma \etac$
and $\piz X_i$ selection criteria are estimated in each $M(X_i)$ mass
bin using phase space distributed $\psp \to \piz X_i$ MC events.
Combining this relative efficiency with the number of $\psp \to
\piz X_i$ signal events in every $M(X_i)$ bin, we estimate the $\piz
X_i$ events that pass the $\gamma\etac$ selection.  We also
examine the efficiencies of $\piz X_i$ events generated with
different dynamics, and the change is negligible.

Other potential $\psp$ decay backgrounds are investigated using $1.06\times10^8$
inclusive MC events where $\piz X_i$ events have been excluded.  We find no other
dominant background processes, but do find dozens of decay modes that each
makes small additional contributions to the background.  These decays typically have
additional or fewer photons in their final states. The sum of these background events is
used to estimate the contribution from other $\psp$ decays. Backgrounds from the $e^+e^-
\to q\bar{q}$ continuum process are studied using a data sample taken at
$\sqrt{s}=3.65~\gev$. Continuum backgrounds are found to be small and uniformly
distributed in $M(X_i)$. There is also an irreducible non-resonant background, $\psp
\to \gamma X_i$, that has the same final state as signal events. A non-resonant component
is included in the fit to the $\etac$ invariant mass.

\begin{figure*}[htbp]
  \psfig{file=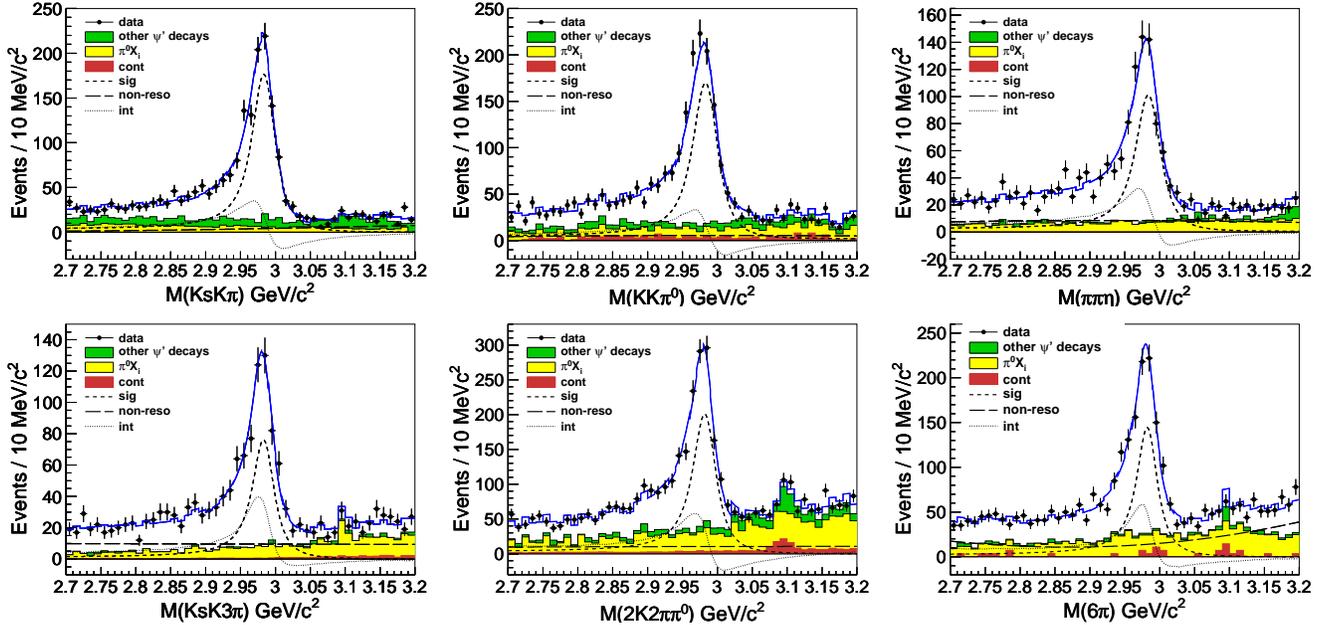}
  \caption{The $M(X_i)$ invariant mass distributions for the decays $\kskp$, $\kkpiz$,
  $\etapp$, $\kskppp$, $\kkpppiz$ and $\pppppp$, respectively, with the fit results (for
  the constructive solution) superimposed. Points are data and the various curves are
  the total fit results. Signals are shown as short-dashed lines, the non-resonant
  components as long-dashed lines, and the interference between them as dotted lines.
  Shaded histograms are (in red/yellow/green) for (continuum/$\pi^0 X_i$/other $\psp$
  decays) backgrounds.  The continuum backgrounds for $\kskp$ and $\etapp$ decays are
  negligible.}
  \label{fig:metac}
\end{figure*}

Figure~\ref{fig:metac} shows the $\etac$ invariant mass
distributions for selected $\etac$ candidates, together with the
estimated $\piz X_i$ backgrounds, the continuum backgrounds
normalized by luminosity, and other $\psp$ decay
backgrounds estimated from the inclusive MC sample. A clear $\etac$
signal is evident in every decay mode. We note that all of the $\etac$
signals have an obviously asymmetric shape: there is a long tail on the low-mass
side; while on the high-mass side, the  signal drops rapidly and the data
dips below the expected level of the smooth background. This behavior of the
signal suggests possible interference with the non-resonant 
$\gamma X_i$ amplitude. In this analysis, we assume
100\% of the non-resonant amplitude interferes with the $\etac$.

The solid curves in Fig.~\ref{fig:metac} show the results of an unbinned simultaneous
maximum likelihood fit in the range from $2.7$ to $3.2~\mathrm{GeV/c^2}$ with three
components: signal, non-resonant background, and a combined background consisting of $\piz
X_i$ decays, continuum, and other $\psp$ decays.  The signal is described by a
Breit-Wigner function~($BW$) convolved with a resolution function. The non-resonant amplitude is
real, and is described by an expansion to second order in Chebychev polynomials defined and normalized over the
fitting range. The combined background is fixed at its expected intensity, as described
earlier. The fitting probability density function (PDF) as a function of mass~($m$) reads:
$$
F(m) = \sigma \otimes \left[\epsilon(m)\left|e^{i\phi}E_\gamma^{7/2} {\cal S}(m) +\alpha{\cal N}(m)\right|^2\right]+ {\cal B}(m)
$$ where ${\cal S}(m)$, ${\cal N}(m)$ and ${\cal B}(m)$ are the signal, the non-resonant
$\gamma X_i$ component, and the combined background, respectively; $E_\gamma$ is photon
energy; $\sigma$ is the experimental resolution and $\epsilon(m)$ is the mass-dependent
efficiency. The $E_\gamma^7$ multiplying $|{\cal S}(m)|^2$ reflects the expected
energy dependence of the hindered-$M1$ transition~\cite{Brambilla:2005zw}, which partially
contributes to the $\eta_c$ low mass tail as well as the interference effect. The
interference phase, $\phi$, and the strength of the non-resonant component, $\alpha$, are
allowed to vary in the fit.

The mass-dependent efficiencies are determined from phase space
distributed MC simulations of the $\etac$ decays.
Efficiencies obtained from MC samples that include intermediate states change the
resulting mass and width by negligible amounts. MC studies indicate
that the resolution is almost constant over the fitting range.
Thus, a mass-independent resolution is used in the fit. The
detector resolution is primarily determined by MC simulation for
each $\etac$ decay mode. The consistency between the data and MC
simulation is checked by the decay $\psp\to \gamma\gamma \jpsi$,
where the $\jpsi$ decays into the same final states as the $\etac$. We
use a smearing Gaussian function to describe possible discrepancies
between data and MC simulations. By fitting the MC-determined $\jpsi$ shape
convolved by a smearing Gaussian function to the data, we determine the
parameters of the Gaussian function.  Due to the different kinematics, the
parameters are slightly different for each mode.

In the simultaneous fit, the $\etac$ mass and width are constrained to be the
same for all the decay modes but still free parameters; the two Chebyshev polynomial coefficients
and the factor $\alpha$ are also allowed to float.  Two solutions for the relative phase
are found for each decay mode, one corresponds to constructive and the other destructive
interference between the two amplitudes at the $\eta_c$ peak. Regardless of which solution
we take, the mass, width of the $\etac$ and the overall fit quality are always
unchanged~\cite{Zhu:2011ha}.  The mass is $M = 2984.3\pm 0.6~\mevcc$, and width $\Gamma =
32.0\pm 1.2~\mev$.  The goodness-of-fit $\chi^2/ndf=283.4/274$, which indicates a
reasonable fit. The solutions for relative phase of each mode are listed in 
Table~\ref{table}.

\begin{table}[htbp]
  \caption{Solutions of relative phase (in unit of radian) of each decay mode.}
  \begin{center}
    \begin{tabular}{ l  c  c}\hline
      mode &~~ constructive ~~& ~~ destructive~~ \\ \hline
$\kskp$   &  $ 2.94 \pm 0.27 $ &   $ 3.75 \pm 0.26 $ \\ \hline
$\kkpiz$   &  $ 2.63 \pm 0.21 $ &   $ 3.96 \pm 0.19 $ \\ \hline
$\etapp$   &  $ 2.41 \pm 0.13 $ &   $ 4.28 \pm 0.09 $ \\ \hline
$\kskppp$   &  $ 2.16 \pm 0.11 $ &   $ 4.46 \pm 0.07 $ \\ \hline
$\kkpppiz$ &  $ 2.73 \pm 0.19 $ &   $ 4.00 \pm 0.16 $ \\ \hline
$\pppppp$      &  $ 2.28 \pm 0.10 $ &   $ 4.43 \pm 0.06 $ \\ \hline
\end{tabular}
\label{table}
\end{center}
\end{table}

However, without the interference term, the fit would miss some data points, especially
where the symmetric shape of a Breit-Wigner function is deformed, and the goodness-of-fit is
$\chi^2/ndf=426.6/280$. The statistical significance of the
interference,
calculated based on the differences of likelihood and degrees
of freedom between fits with and without interference,
is of order 15$\sigma$.


The systematic uncertainties of the $\etac$ mass and width  mainly
come from the background estimation, the mass scale and resolution,
the shape of the non-resonant component, the fitting range, and the
efficiency.

In the fit, the $\piz X_i$ background is fixed at its expected
intensity, so the statistical uncertainty of the observed $\piz X_i$ events
introduces a systematic error. To estimate this uncertainty, we vary the
number of events in each bin by assuming Gaussian variations from
the expected value. We repeat this procedure a thousand times,
and take the standard deviation of the resulting mass, width, and phases as
systematic errors. We also use different dynamics in generating the
$\piz X_i$ events (with the same final state, but different
intermediate states) for the efficiency correction, and find the
differences in resulting mass and width are small. We take
$0.24~\mevcc$ in mass and $0.44~\mev$ in width as the
systematic errors for the $\piz X_i$ background estimation.

We assign a $0.07~\mevcc$ ($0.06~\mev$) error in mass (width) for the
non-resonant component shape that is obtained by changing the
polynomial order. Also we include an additional non-interfering
component, which is represented by a 2nd-order polynomial with free
strength and shape parameters. The changes in the resulting $\eta_c$
mass and width are 0.10~MeV/$c^2$ and 0.02~MeV, respectively; and the 
fraction of this component to total non-resonant rate varies from
0 to 25\% depending on decay mode. These variations are included in
systematic errors. 

The systematic error from the uncertainty in the other $\psp$ decay
backgrounds is estimated by floating the magnitude and changing the
shape of this component to a 2nd-order polynomial with free
parameters. The changes, $0.05~\mevcc$ in mass and $0.06~\mev$ in
width, are taken as systematic errors.

The consistency of the mass scale and resolution between data and MC simulations is checked with the
decay $\psp\to \gamma\gamma \jpsi$, and possible discrepancies are described by a
smearing Gaussian distribution, where a non-zero mean value indicates a mass offset, and a
non-zero $\sigma$ represents difference between the data and MC mass
resolutions $(\sigma_{\rm data}^2-\sigma_{\rm MC}^2)^{1/2}$. A typical mass shift is
about $-1.0$~$\mevcc$  and resolution smear is $\sim 3.0$ $~\mev$.
Another possible bias is the difference between input and the value
after event-reconstruction and selection. This is small for both the
mass shift ($<0.3~\mev$) and resolution smear. Both of these are added in
the smearing Gaussian distribution.
By varying the parameters of the smearing Gaussian
distribution from the expected value, we estimate the uncertainties. From a large number
of tests, the standard deviation of the resulting mass (width), $0.38~\mevcc$ ($0.27~\mev$), is
taken as a systematic error in mass (width) for the mass scale uncertainty. A
$0.35~\mevcc$ ($0.60~\mev$) systematic error in mass (width) is assigned due to the mass
resolution uncertainty.

The systematic error due to the fitting range is estimated by varying the
lower-end between 2.6 and 2.8~$\gevcc$ and the higher-end between 3.1 and 3.3~$\gevcc$.
The changes, $0.05~\mevcc$ in mass and $0.07~\mev$ in width, are assigned as systematic errors. A
mass-dependent efficiency is used in the fit. By removing the efficiency correction from
the fitting PDF, the changes, which are $0.05~\mevcc$ in mass and $0.06~\mev$ in width,
are taken as systematic errors. The stability of the simultaneous fit program is checked
by repeating the fit a thousand times with random initialization; the standard deviation
of mass and width, $0.14~\mevcc$ and $0.66~\mev$, respectively, are taken as systematic
errors.

We assume all these sources are independent and take their sum in
quadrature as the total systematic error. 
We obtain the $\etac$ mass and width to be
$$ M =2984.3\pm 0.6\pm 0.6~\mevcc,$$
$$\Gamma = 32.0\pm 1.2\pm 1.0~\mev.$$
Here (and elsewhere) the first errors are statistical and the second are
systematic.

The relative phases for constructive interference or destructive
interference from each mode are consistent with each other within
$3\sigma$, which may suggest a common phase in all the modes under
study. A fit with a common phase ({\it i.e.} the phases are constrained to
be the same) describes the data well, with a $\chi^2 / ndf=303.2/279$.
Comparing to the fit with separately varying phases for each mode, we
find the statistical significance for the case of five distinct phases to be
3.1$\sigma$. This fit yields 
$M    = 2983.9\pm  0.6\pm 0.6~\mevcc$, 
$\Gamma = 31.3\pm  1.2\pm 0.9~\mev$, and 
$\phi =   2.40\pm 0.07\pm 0.47$~rad (constructive) or 
$\phi =   4.19\pm 0.03\pm 0.47$~rad (destructive).
The physics behind this possible common phase is yet to
be understood.


In summary, we measure the $\etac$ mass and width via $\psp\to \gamma
\etac$ 
by assuming all radiative non-resonant events interfere with the $\etac$.  These results
are so far the most precise single measurement of the mass and width of
$\eta_c$~\cite{PDG}. For the first time, interference between the $\etac$ and the
non-resonant amplitudes around the $\etac$ mass is considered; given the assumptions of
our fit, the significance of the interference is of order $15\sigma$.
We note that this interference affects the $\etac$ mass and width significantly, and may
have impacted all of the previous measurements of the $\etac$ mass and
width that used radiative transitions.  Our results are consistent with
those from photon-photon fusion and $B$ decays~\cite{Asner:2003wv, Aubert:2003pt,
Uehara:2007vb, belle2011}; this may partly clarify the discrepancy
puzzle discussed above. The changes of the $\etac$ mass and width may
also have an impact on the expected $\etacp$ mass and width, and will
modify the parameters used in charmonium potential models, where the
$\etac$ mass is one of the input parameters. From this measurement,
we determine the hyperfine mass splitting to be
$\Delta M_{hf}(1S)_{c\bar{c}} \equiv M(\jpsi)-M(\etac) =112.6\pm 0.8~\mevcc$,
which agrees well with recent lattice computations~\cite{Burch:2009az,
Levkova:2010ft, Kawanai:2011jt} as well as quark model predictions~\cite{splitting},
and sheds light on spin-dependent interactions in quarkonium states.

The BESIII collaboration thanks the staff of BEPCII and the IHEP computing center for their strong efforts. This work is supported in part by the Ministry of Science and Technology of China under Contract No. 2009CB825200; National Natural Science Foundation of China (NSFC) under Contracts Nos. 10625524, 10821063, 10825524, 10835001, 10935007; Joint Funds of the National Natural Science Foundation of China under Contract No. 11079008; the Chinese Academy of Sciences (CAS) Large-Scale Scientific Facility Program; CAS under Contracts Nos. KJCX2-YW-N29, KJCX2-YW-N45; 100 Talents Program of CAS; Istituto Nazionale di Fisica Nucleare, Italy; Siberian Branch of Russian Academy of Science, joint project No 32 with CAS; U. S. Department of Energy under Contracts Nos. DE-FG02-04ER41291, DE-FG02-91ER40682, DE-FG02-94ER40823; U.S. National Science Foundation; University of Groningen (RuG) and the Helmholtzzentrum fuer Schwerionenforschung GmbH (GSI), Darmstadt; WCU Program of National Research Foundation of Korea under Contract No. R32-2008-000-10155-0.
This paper is also supported by the NSFC under Contract Nos. 10979038,
  10875113, 10847001, 11005115; Innovation Project of Youth Foundation of Institute of
  High Energy Physics under Contract No. H95461B0U2.


\begin{thebibliography}{**}

\bibitem{many} See for example: E.~S.~Swanson,
  Phys.\ Rept.\  {\bf 429}, 243 (2006);
  E.~Klempt and A.~Zaitsev, Phys. Rept. {\bf 454}, 1 (2007);
  S.~Godfrey and S.~L.~Olsen, Ann. Rev. Nucl. Part.
  Sci. {\bf 58}, 51 (2008); N.~Brambilla {\it et al.}, Eur.\
  Phys.\ J. C {\bf 71}, 1534 (2011).


\bibitem{PDG} K.~Nakamura {\it et al.} [Particle Data Group],
  J.\ Phys.\ G {\bf 37}, 075021 (2010).

\bibitem{Baltrusaitis:1985mr}
  R.~M.~Baltrusaitis {\it et al.}  [Mark-III Collaboration],
  Phys.\ Rev.\  D {\bf 33}, 629 (1986).

\bibitem{Bai:2003et}
  J.~Z.~Bai {\it et al.}  [BES Collaboration],
  Phys.\ Lett.\  B {\bf 555}, 174 (2003)
  [arXiv:hep-ex/0301004].

\bibitem{Asner:2003wv}
  D.~M.~Asner {\it et al.}  [CLEO Collaboration],
  Phys.\ Rev.\ Lett.\  {\bf 92}, 142001 (2004)
  [arXiv:hep-ex/0312058].

\bibitem{Aubert:2003pt}
  B.~Aubert {\it et al.}  [BABAR Collaboration],
  Phys.\ Rev.\ Lett.\  {\bf 92}, 142002 (2004)
  [arXiv:hep-ex/0311038].

\bibitem{Uehara:2007vb}
  S.~Uehara {\it et al.}  [Belle Collaboration],
  Eur.\ Phys.\ J.\  C {\bf 53}, 1 (2008)
  [arXiv:0706.3955 [hep-ex]].

\bibitem{belle2011}
  A.~Vinokurova {\it et al.} [Belle Collaboration],
  Parameters,''
  Phys. Lett B{\bf 706}, 139, (2011)
  [arXiv:1105.0978 [hep-ex]].

\bibitem{recent:2008fb}
  R.~E.~Mitchell {\it et al.}  [CLEO Collaboration],
  Phys.\ Rev.\ Lett.\  {\bf 102}, 011801 (2009)
  [arXiv:0805.0252 [hep-ex]].


\bibitem{npsp} 
  M.~Ablikim {\it et al.} [BESIII Collaboration],
  Phys.\ Rev.\ D {\bf 81}, 052005 (2010).

\bibitem{bes3}
  M.~Ablikim {\it et al.} [BESIII Collaboration],
  Nucl.\ Instrum.\ Meth.\ A {\bf 614}, 345 (2010).

\bibitem{KKMC}
S.~Jadach, B.~F.~L.~ Ward and Z.~Was, Comp. Phys. Commu. {\bf 130},
260 (2000); Phys. Rev. D {\bf 63}, 113009 (2001).

\bibitem{EvtGen}
 R.~G.~Ping {\it et al.}, Chinese Physics C {\bf 32}, 243 (2008);
http://www.slac.stanford.edu/\~lange/EvtGen/

\bibitem{Chen:2000tv}
  J.~C.~Chen, G.~S.~Huang, X.~R.~Qi, D.~H.~Zhang and Y.~S.~Zhu,  
  Phys.\ Rev.\  D {\bf 62}, 034003 (2000).

\bibitem{geant4} S.~Agostinelli {\it et al.}
[{\sc geant4} Collaboration], Nucl.\ Instrum.\ Meth.\ A {\bf 506},
250 (2003).




\bibitem{Brambilla:2005zw} 
  N.~Brambilla, Y.~Jia and A.~Vairo,
  Phys.\ Rev.\ D {\bf 73}, 054005 (2006)
  [hep-ph/0512369].

\bibitem{Zhu:2011ha}

  K.~Zhu, X.~H.~Mo, C.~Z.~Yuan and P.~Wang,
  Int.\ J.\ Mod.\ Phys.\  A {\bf 26}, 4511 (2011)
  [arXiv:1108.2760 [hep-ex]].


\bibitem{Burch:2009az}
  T.~Burch {\it et al.},
  Phys.\ Rev.\  D {\bf 81}, 034508 (2010)
  [arXiv:0912.2701 [hep-lat]].


\bibitem{Levkova:2010ft}
  L.~Levkova and C.~DeTar,
  Phys.\ Rev.\  D {\bf 83}, 074504 (2011)
  [arXiv:1012.1837 [hep-lat]].

\bibitem{Kawanai:2011jt}
  T.~Kawanai and S.~Sasaki,
  [arXiv:1110.0888 [hep-lat]], [Phys. Rev. D (to be published)]

\bibitem{splitting} K.~K.~Seth, [arXiv:0912.2776v1 [hep-ex]].


\end{thebibliography}
\end{document}